# A Novel Framework for Characterization of Tumor-Immune Spatial Relationships in Tumor Microenvironment

Mahmudul Hasan, Jakub R. Kaczmarzyk, David Paredes, Lyanne Oblein, Jaymie Oentoro, Shahira Abousamra, Michael Horowitz, Dimitris Samaras, Chao Chen, Tahsin Kurc, Kenneth R. Shroyer*, Joel Saltz*

Stony Brook University, Stony Brook, NY, USA
Kenneth.Shroyer@stonybrookmedicine.edu,
joel.saltz@stonybrookmedicine.edu

**Abstract.** Understanding the impact of tumor biology on the composition of nearby cells often requires characterizing the impact of biologically distinct tumor regions. Biomarkers have been developed to label biologically distinct tumor regions, but challenges arise because of differences in the spatial extent and distribution of differentially labeled regions. In this work, we present a framework for systematically investigating the impact of distinct tumor regions on cells near the tumor borders, accounting their cross spatial distributions. We apply the framework to multiplex immunohistochemistry (mIHC) studies of pancreatic cancer and show its efficacy in demonstrating how biologically different tumor regions impact the immune response in the tumor microenvironment. Furthermore, we show that the proposed framework can be extended to large-scale whole slide image analysis.

## 1 Introduction

The analysis of spatial relationships among tumor infiltrating immune cell types relative to cancer biomarker expression can be used to develop more effective approaches for immunotherapeutic intervention in a wide range of solid tumors, including pancreatic ductal adenocarcinoma (PDAC) [1,2]. The use of high-throughput slide scanning and innovative machine learning-based algorithms to relate cancer biomarker expression to the inflammatory tumor microenvironment (TME), using either state-of-the-art brightfield, multiplexed immunohistochemical (mIHC) or multiplexed immunofluorescent (mIF) protocols can provide automated, unbiased, and reproducible analyses across hundreds of specimens. In our work, we seek to develop methods that characterize the effects of differential tumor regions on immune cells. We focus on the analysis of biologically relevant subclasses of T cells and macrophages, using mIHC, to generate data on immune cell quantity and spatial relationships to tumor cell, relative to pan-cytokeratin, a marker of tumor epithelial cells, and to keratin 17 (K17), a biomarker of the most biologically aggressive form of PDAC [3,4,5]. The mIHC-based analysis has provided unique insight into the spatial relationships among cells within the complex TME, including infiltrating immune cells, cancer cells, and stromal cells [6]. An earlier work has reported a deep learning approach and a multiparameter quantification strategy to enable comprehensive phenotyping of immune complexity [7]. Building upon the deep learning approach, we now report a modified method to carry out a comparative analysis between immune cells and biomarker-positive tumor cells in order to further develop an understanding of the TME. The primary innovation of our work is the introduction of a novel, image-based scoring system of cancer biomarkers. The scoring system computes a metric, called "influence score", by a spatial analysis of deep learning-based segmentation and classification results, to measure and compare the influence of tumor regions on different types to immune cells. The proposed metric is carefully designed with biological rationale. We demonstrate how the influence score can be applied to multiplex IHC PDAC data to analyze the effect of K17 on different types of immune cells.

## 2 Methodology

Brightfield mIHC of tissue sections was performed using an Ultra Autostainer (Roche/Ventana, Oro Valley, AZ). Antibodies for CD4, CD8, CD16, CD163, pancytokeratin, and K17 were provided by Roche/Ventana. mIHC staining was performed using horseradish peroxidase (HRP) and alkaline phosphatase (AP)-based protocols with different colored chromogens (Red: CD4, Purple: CD8, Yellow: CD163, Green: CD16, Teal: pancytokeratin, and Brown: K17) to enable multispectral imaging of diverse immune cell populations within the cancer microenvironment. A deep learning analysis workflow was used to detect and classify cells in whole slide images (WSIs) of mIHC stained tissue specimens. The WSIs were generated at a resolution of 0.346 μm per pixel using an Olympus VS120 digital microscope (Olympus, Tokyo, Japan). The analysis workflow is illustrated in Fig. 1.

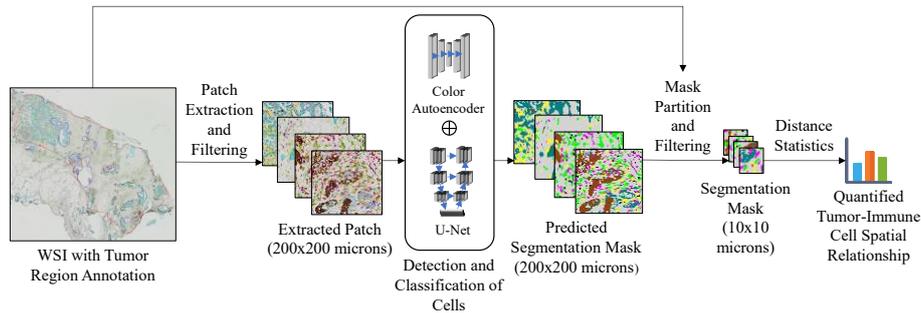

**Fig. 1.** Multiplex Immunohistochemical Whole Slide Image Analysis Pipeline

The workflow consists of two deep learning models combined in an ensemble configuration [7]. Briefly, pixel-wise predictions from a color auto-encoder (ColorAE) model [8] and a U-Net model [9] are combined to create multi-class masks. A labeled pixel in a mask indicates to which type of cell (e.g., CD4, CD8 lymphocyte) the pixel belongs. The masks are further analyzed as described in the rest of this section. The notations used in the following sections are described in Table S1 (Supplementary Material).

### 2.1 Cell Detection and Classification

The ensemble of ColorAE and U-Net was developed in [7] for detection and classification of cells in mIHC images. Each model is trained separately, and predictions from each model are combined in the inference phase to create multi-class masks. The predictions can be combined in four ways (union, intersection, Union anchor AE, Union anchor U-Net). We used the Union anchor U-Net configuration, which we experimentally validated to be the best performing configuration for our dataset.

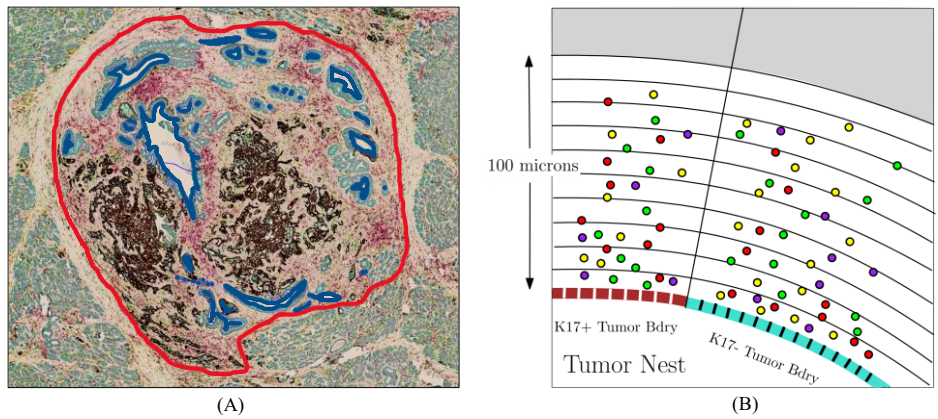

(A)  (B)

**Fig. 2.** (A) Tumor and non-tumor region annotations overlaid on the multiplex IHC slide (B) Schematic diagram of tumor nest with K17-positive and K17-negative boundary and immune cells distributed within 100 μm

The multiplex semantic segmentation ensemble is applied on patches of size 580 x 580 pixels extracted from WSIs at 0.346 μm/pixel resolution. The patches are extracted from regions of interest, which are primarily composed of tumor tissue and selected manually by pathologists. The pathologists also annotate sub-regions to be excluded from the analysis (e.g., empty spaces, artifacts, etc.) and mark them as 'non-tumor', see Fig. 2A. The patches (P) from the WSI are filtered to keep only those intersecting with the selected tumor regions. Multi-class masks are generated from the final set of patches by the ensemble deep learning method.

## 2.2    Expressing WSIs in terms of K17

We partition each mask into a 20 x 20 grid. We divide the mask into patches of 29 x 29 pixels (~ 10 x 10 microns). We filter the mask patches keeping only those in the regions of interest, and furthermore, excluding those intersecting with the marked non-tumor regions. Using the set of mask patches, we compute a K17 score, which quantitatively characterizes a WSI in terms of tumor percentage, as follows:

$$K17\ score = \frac{\text{Area}_{tc=K17+}}{\text{Area}_{tc=K17+} + \text{Area}_{tc=K17-}} \qquad (1)$$

In equation (1) the numerator stands for area of K17 positive (K17+) regions in a WSI and the denominator represents the area of K17 positive and K17 negative (K17-) regions. If we denote 29 x 29 pixels mask as $\{m_1, m_2, \ldots, m_N\} \in m$ and pixel in a tumor cell (i.e. K17+ and K17- segmentation masks) from mask $m$ as $px_{m_i}^{tc}$ (where, $\{K17+, K17-\} \in tc$), we can calculate the area of tumor by $Area_{tc} = \sum_{i=1}^{i=|m|} px_{m_i}^{tc}$

## 2.3    Quantification of Tumor-Immune Cell Spatial Relationship

We propose a scoring system (illustrated in Fig. 3.) to quantify the influence of tumor microenvironment on the distribution of different categories of immune cells. For this purpose, we introduce a novel metric called tumor/stroma influence score denoted by $I_M^{ic}$ by calculated by following equation:

$$I_M^{ic} = \frac{Cell\ Count_M^{ic}}{Influence\ Area_M} \qquad (2)$$

The equation represents the approximate count of each immune cell (numerator) normalized by total influence area (denominator). In the equation, $ic$ represents immune cells (CD4, CD8, CD16 and CD163) and $M$ represents the tumor boundary explained later in this section. To calculate the influence area, we define tumor nest area using 29x29 pixels mask $m$ containing only tumor cell denoted as $tc$. We call $m$ a tumor nest mask ($m_{tc}$), if the percentage of tumor cell area within that mask area is more than 35% (decided by expert pathologists). We further subclassify $m_{tc}$ into $m_{tc=K17+}$ mask if more than 40% area of the tumor cells contain K17+ (decided by expert pathologists) else it is considered to be $m_{tc=K17-}$. The $m_{tc=K17+}$ and $m_{tc=K17-}$ are visually illustrated in Fig. 3(B). We accumulate the tumor nest masks, define tumor nests with K17+ and K17- regions and get a rectangular boundary $M$ containing two distinct parts denoted by marker $M^+$ and marker $M^-$. The tumor nest boundaries with $M^+$ (brown colored lines) and $M^-$ (teal-colored lines) are depicted in Fig. 3(C). We are representing a tumor nest having two distinct regions with boundaries having two distinct parts so that we can compute the approximate relationship between tumor and immune cells in an experimentally achievable time. Next, we define a partitioned influence area having K17+ and K17- influenced region. To calculate the influence area, we consider the region that falls outside the tumor nest and within the vicinity of 100μm tumor microenvironment from a tumor nest boundary.

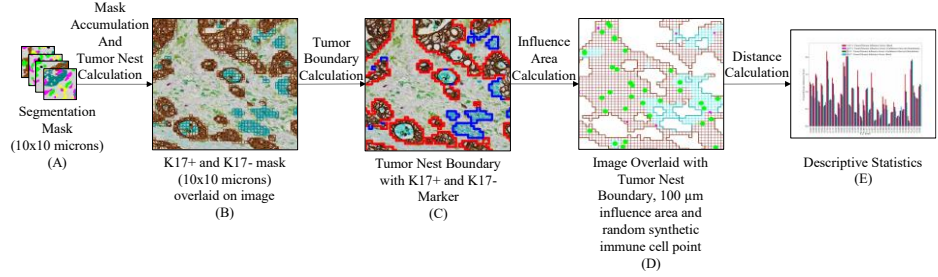

**Fig. 3.** Mechanism of Distance Based Tumor-Immune Spatial Relationships Calculation

A conceptual diagram is shown in Fig. 2(B). For this purpose, we make use of the 29x29 pixels mask $m$ that is not classified as tumor nest mask $m_{tc}$ denoted by, $m_{ntc} = m - m_{tc}$. This $m_{ntc}$ represents the discretized non tumor nest space within the tumor region. Considering center point of a single mask $m_{ntc}^{(i)}$ with index $(i)$ as $C_{m_{ntc}}^{(i)}$ (where, $i : [1, |m_{ntc}|] \in R$) and boundary $M$, we classify each $m_{ntc}$ as $M^+$ or $M^-$ influenced mask. We denote a single mask $m_{ntc}^{(i)}$ as K17+ influenced mask denoted by $m_{M^+}^{(i)}$ if the euclidean distance between center point of that mask and the nearest $M^+$ boundary is less than the nearest $M^-$ boundary and within 100μm distance from the $M^+$. Otherwise, if the distance between center point and the nearest $M^-$ boundary is less than the nearest $M^+$ boundary and within 100μm distance from the $M^-$ then it is considered K17- influenced mask denoted by $m_{M^-}^{(i)}$. This can be expressed by the following equation:

$$m_M^{(i)} = argmin_{M^+, M^-} \left( \left\| C_{m_{ntc}}^{(i)} - M^+ \right\|_2^2 \times mpp, \left\| C_{m_{ntc}}^{(i)} - M^- \right\|_2^2 \times mpp \right) \\ \leq 100 \text{ μm} \tag{3}$$

This method partitions the influence space into two regions: $M^+$ influenced region (brown colored squares) and $M^-$ influenced region (teal colored squares) depicted in Fig. 3(D). In a similar manner, considering $px_{m_{ntc}}^{ic(i,j)}$ (where, $j : [1, 29 \times 29] \in R$) as a single pixel with index $(j)$ of immune cells in a single mask, $m_{ntc}^{(i)}$ with index $(i)$, we can get K17+ and K17- influenced pixels which is expressed by the following equation:

$$px_M^{ic(i,j)} = argmin_{M^+, M^-} \left( \left\| px_{m_{ntc}}^{ic(i,j)} - M^+ \right\|_2^2 \times mpp, \left\| px_{m_{ntc}}^{ic(i,j)} - M^+ \right\|_2^2 \times mpp \right) \\ \leq 100 \text{ μm} \tag{4}$$

Based on equation (3), we accumulate all influenced masks to get the total influenced area for K17+ and K17- by the following equation:

$$Influence\ Area_M = \sum_{i=1}^{i=|m_{ntc}|} m_M^{(i)} \qquad (5)$$

With same technique we can calculate the total influenced immune cell area expressed by following equation:

$$Cell\ Area_M^{ic} = \sum_{i=1}^{i=|m_{ntc}|} \sum_{j=1}^{j=29\times29} px_M^{ic\,(i,j)} \qquad (6)$$

We normalize the influenced cell area by average immune cell area to get an approximate influenced cell count $Cell\ Count_M^{ic}$. We also use this influenced cell count for evaluation purpose using random cell distribution (see Section 5). Our expert pathologists approximated the average lymphocyte (CD4, CD8) to be a circle with an approximate radius of 4 microns and macrophage (CD16, CD163) with a radius of 8 microns. Therefore, the area of the average lymphocyte is $\pi r^2 = 3.1416 \times 4^2 = 50\ \mu m^2$ or $144\ pixel^2$ (approx.), and the area of the average macrophage is $201\ \mu m^2$ or $580\ pixel^2$ (approx.) for our calculation. Finally, we plug in equation (5) and $Cell\ Count_M^{ic}$ and thus get our influence score for each immune cell.

## 3    Dataset and Model Training

The training and validation datasets for the deep learning models were generated from 23 WSIs. Expert pathologists manually put labeled dots into the vicinity of each cell over 1000x1000 pixel tiles extracted from these 23 WSIs.

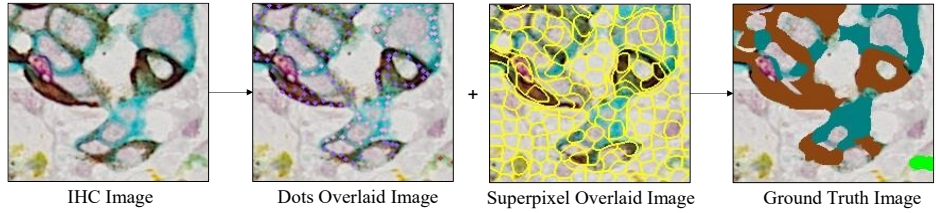

IHC Image          Dots Overlaid Image          Superpixel Overlaid Image          Ground Truth Image

**Fig. 4.** Annotation Process for Segmentation

**Table 1.** Dataset Description

|  |  |  |  | **Train/Validation/Test** |
|---|---|---|---|---|
| Number of WSIs | 23 | Number of non-overlapping patches (1000x1000) |  | 57/4/6 |
| Total Number of non-overlapping patches | 67 | Number of overlapping patches (400x400) |  | 912/64/- |

To speed up the manual annotation process, we generated superpixel masks using simple linear iterative clustering (SLIC) [10] with 6000 segments. The dot labels are expanded to pixel-wise label maps by superimposing the input image superpixel map and assigning a dot's label to all the pixels in the overlapping superpixel. We subdivided the tiles into 16 400400 patches with an overlap of 200 pixels in each dimension to generate ground truth. Figure 4 shows the annotation workflow. The training and validation datasets generated through this process are described in Table 1.

## 4 Model Validation and Experimental Setup

For quantitative evaluation of our detection and classification model we used F1 score illustrated in Table 2. We also performed qualitative analysis (See Fig. S1. in the supplementary material). Based on both quantitative and qualitative results, we found Union Anchor U-Net is the best model to perform detection and classification.

**Table 2.** F1 Score Result of our Detection and Classification Models

| F1 Score ↑ | CD4 | CD8 | CD16 | CD163 | K17+ | K17- |
|---|---|---|---|---|---|---|
| Intersection | 0.80 | 0.89 | 0.77 | 0.90 | 0.91 | 0.97 |
| Union | 0.91 | 0.87 | 0.75 | 0.92 | 0.97 | 0.98 |
| Union Anchor AE | 0.84 | 0.89 | **0.81** | 0.92 | 0.93 | 0.98 |
| Union Anchor U-Net | **0.92** | **0.89** | 0.77 | **0.93** | **0.98** | **0.99** |

We used a dropout rate of 0.3 in the U-Net and the following color concentration thresholds in the colorAE model: 0.7 for K17-positive, 0.1 for K17-negative, 0.1 for CD4, 0.1 for CD8, 0.1 for CD16, and 0.1 for CD163. We carried out our analyses on multiple shared V100-32GB SXM2 GPUs and AMD EPYC "Rome" 7742 64-core CPUs provided by the Extreme Science and Engineering Discovery Environment (XSEDE) [11].

## 5 Results and Conclusions

We have implemented our framework on 40 WSIs to find a pattern between immune cells and tumor biomarkers. Fig. 6 shows our distance analysis result using 25μm instead of 100μm for 40 WSIs. For evaluation we have used 25 homogeneous poisson point process [12, 13, 14] simulations to scatter each type of cell on the environment space randomly. This simulation addresses our null hypothesis that there is a relationship between the tumor boundary and cell distribution. In the figure, the x-axis represents K17 score, and y-axis represents Tumor/Stroma Influence score. Each group of four bars in the graph corresponds with a single WSI. In the figure, we have only shown results for CD163, where we can observe that the real Tumor/Stroma Influence score largely deviates from the simulation confidence band and there is a noisy linear trend with respect to Influence Score. For an ideal case (i.e., if the cells are randomly distributed) we can observe that K17+ and K17- simulation confidence are approximately equivalent to each other. We have calculated the mean absolute deviation (MAD) between 25 simulations mean and real scenarios for both K17+ and K17-, which is tabulated in Table 3. It is clarified from the table that CD8 and CD163 have higher deviation from the randomness compared to CD4 and CD16, implicating CD8 and CD163 have spatial patterns.

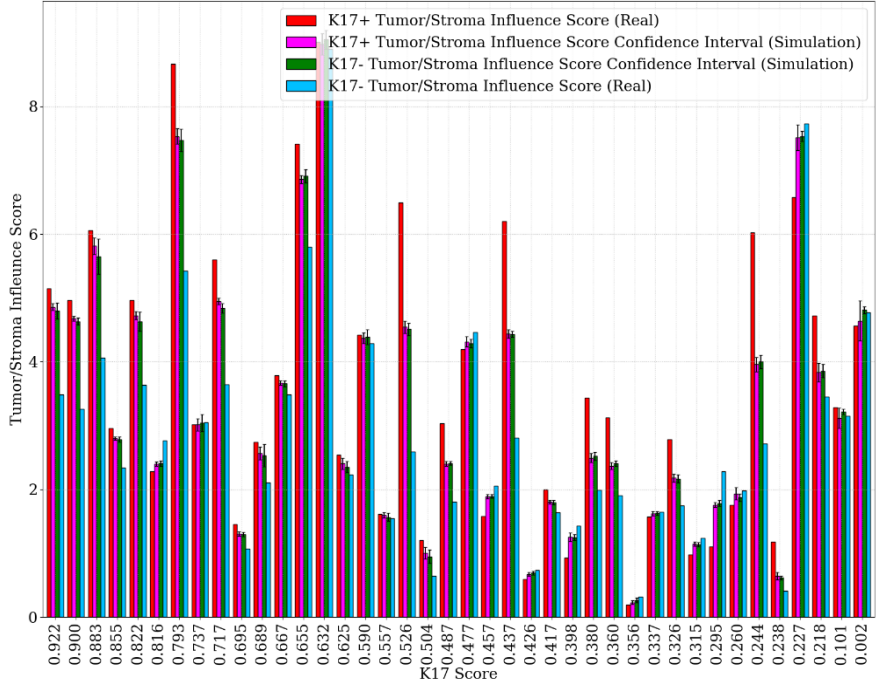

**Fig. 5.** Influence Score vs K17 Score for CD163 in $25\mu m$ Tumor Environment

**Table 3.** Deviation Result Between the Real Scenario and 25 Homogeneous Poisson Point Process Simulation (MAD – Mean Absolute Deviation)

| Immune Cell Type | K17+ MAD | K17- MAD | Total MAD |
|---|---|---|---|
| CD4 | 0.15 | 0.15 | 0.30 |
| CD8 | 0.69 | 0.67 | 1.36 |
| CD16 | 0.18 | 0.15 | 0.34 |
| CD163 | 0.45 | 0.53 | 0.99 |

Though this initial noisy pattern and correlation does not conclude the full story of tumor-immune relationship, this preliminary result is inspiring for us to conduct more investigative research on a larger set of data and subdivide into groups to fully understand the internal mechanism of Tumor-Immune relationship. From a technical perspective, we also intend to use the spatial results as a prior for developing a more advanced deep learning-based framework.

## Acknowledgement

This study was supported in part by a sponsored Research Agreement to KRS from Roche Diagnostics Corporation.

## References

1. Fu, T., Dai, LJ., Wu, SY. et al. Spatial architecture of the immune microenvironment orchestrates tumor immunity and therapeutic response. J Hematol Oncol 14, 98 (2021).
2. Murciano-Goroff, Y.R., Warner, A.B. & Wolchok, J.D. The future of cancer immunotherapy: microenvironment-targeting combinations. Cell Res 30, 507–519 (2020).
3. Roa-Peña, C.V. Leiton, S. Babu, E.A. Vanner, C.-H. Pan, A. Akalin, J. Bandovic, R.A. Moffitt, L.F. Escobar-Hoyos, K.R. Shroyer. Keratin 17 identifies the most lethal molecular subtype of pancreatic cancer. Sci Reports, 9:11239, 2019 PMID: 31375762
4. L. Roa-Peña, S. Babu, C.V. Leiton, M Wu, S. Taboada, A. Akalin, J. Jonathan Buscaglia, L.F. Escobar-Hoyos, K.R. Shroyer. Keratin 17 testing in pancreatic cancer needle aspiration biopsies predicts survival. Cancer Cytopathol. 129:865-873, 2021 PMID 34076963
5. G Baraks, R Tseng, CH Pan, S Kasliwal, CV Leiton, KR Shroyer*, LF Escobar-Hoyos*. Dissecting the Oncogenic Roles of Keratin 17 in the Hallmarks of Cancer. In press, Cancer Research (2022)
6. Sun Z, Nyberg R, Wu Y, Bernard B, Redmond WL (2021) Developing an enhanced 7-color multiplex IHC protocol to dissect immune infiltration in human cancers. PLOS ONE 16(2): e0247238.
7. Fassler, Danielle J., et al. "Deep learning-based image analysis methods for brightfield-acquired multiplex immunohistochemistry images." Diagnostic pathology 15.1 (2020): 1-11.
8. Abousamra, Shahira, et al. "Weakly-Supervised Deep Stain Decomposition For Multiplex IHC Images." IEEE International Symposium on Biomedical Imaging (ISBI), 2020.
9. Ronneberger, Olaf, et al. "U-net: Convolutional networks for biomedical image segmentation." MICCAI, 2015.
10. Achanta, Radhakrishna, et al. "SLIC superpixels compared to state-of-the-art superpixel methods." IEEE transactions on pattern analysis and machine intelligence 34.11 (2012): 2274-2282.
11. John Towns, Timothy Cockerill, Maytal Dahan, Ian Foster, Kelly Gaither, Andrew Grimshaw, Victor Hazlewood, Scott Lathrop, Dave Lifka, Gregory D. Peterson, Ralph Roskies, J. Ray Scott, Nancy Wilkins-Diehr, "XSEDE: Accelerating Scientific Discovery", Computing in Science & Engineering, vol.16, no. 5, pp. 62-74, Sept.-Oct. 2014, doi:10.1109/MCSE.2014.80
12. Miles, Roger E. "On the homogeneous planar Poisson point process." Mathematical Biosciences 6 (1970): 85-127.
13. Jones-Todd, Charlotte M., et al. "Unusual structures inherent in point pattern data predict colon cancer patient survival." arXiv preprint arXiv:1705.05938 (2017).
14. Chervoneva, Inna, et al. "Quantification of spatial tumor heterogeneity in immunohistochemistry staining images." Bioinformatics 37.10 (2021): 1452-1460.